\documentclass{article}

\usepackage{PRIMEarxiv}

\usepackage[utf8]{inputenc} 
\usepackage[T1]{fontenc}    
\usepackage{hyperref}       
\usepackage{url}            
\usepackage{booktabs}       
\usepackage{amsfonts}       
\usepackage{nicefrac}       
\usepackage{microtype}      
\usepackage{lipsum}
\usepackage{fancyhdr}       
\usepackage{graphicx}       
\graphicspath{{media/}}     
\usepackage{tabularx}
\usepackage{multirow}
\usepackage{makecell}
\usepackage{caption}
\usepackage{subcaption}
\usepackage{multirow}
\usepackage{colortbl}
\usepackage{float}
\usepackage{amsmath}
\def\xx{\mathbf{x}}
\def\ff{\mathbf{f}}
\def\bb{\mathbf{b}}

\def\uu{\mathbf{u}}

\def\wn{\mathbf{\tilde{w}}}
\def\wni{\tilde{w}}

\def\bb{\mathbf{b}}

\DeclareMathOperator{\argmax}{arg\,max}

\def\R{{\rm I\!R}}   

\def\cc{\mathbf{s}}
\def\ww{\mathbf{w}}

\def\xgamma{\boldsymbol{\gamma}}
\def\gamman{\tilde{{\boldsymbol{\gamma}}}}
\def\gammani{\tilde{\gamma}}

\title{Polyhedral Conic Classifier for CTR Prediction}

\author{
 Beyza Turkmen \thanks{These authors contributed equally to this work.} \\
  AI Enablement\\
  Huawei Türkiye R\&D Center\\
  Istanbul, Turkey \\
  \texttt{beyza.turkmen2@huawei-partners.com} \\
   \And
 Ramazan Tarik Turksoy \footnotemark[1] \\
  AI Enablement\\
  Huawei Türkiye R\&D Center\\
  Istanbul, Turkey \\
  \texttt{ramazan.tarik.turksoy1@huawei.com}
  \And
Hasan Saribas \\
  AI Enablement\\
  Huawei Türkiye R\&D Center\\
  Istanbul, Turkey \\
  \texttt{hasan.saribas2@huawei-partners.com}
  \And
 Hakan Cevikalp \\
 Electrical and Electronics Engineering Department \\
  Eskisehir Osmangazi University\\
  Eskisehir, Turkey \\
  \texttt{hakan.cevikalp@gmail.com}
}

\begin{document}

\maketitle

\begin{abstract}
This paper introduces a novel approach for click-through rate (CTR) prediction within industrial recommender systems, addressing the inherent challenges of numerical imbalance and geometric asymmetry. These challenges stem from imbalanced datasets, where positive (click) instances occur less frequently than negatives (non-clicks), and geometrically asymmetric distributions, where positive samples exhibit visually coherent patterns while negatives demonstrate greater diversity. To address these challenges, we have used a deep neural network classifier that uses the polyhedral conic functions. This classifier is similar to the one-class classifiers in spirit and it returns compact polyhedral acceptance regions to separate the positive class samples from the negative samples that have diverse distributions. Extensive experiments have been conducted to test the proposed approach using state-of-the-art (SOTA) CTR prediction models on four public datasets, namely Criteo, Avazu, MovieLens and Frappe. The experimental evaluations highlight the superiority of our proposed approach over Binary Cross Entropy (BCE) Loss, which is widely used in CTR prediction tasks.
\end{abstract}


\section{Introduction}

Recommender systems aim to recommend products to users which the users would show interest such as watching, clicking or purchasing. CTR prediction is an important task of recommender systems for online advertising, and CTR prediction performance directly affects user experience and company revenue. Therefore, improving CTR prediction has been a significant focus for both academia and industry \cite{deeplight, fibinet++}.

Deep learning has emerged as a significant driver in the advancement of recommender systems \cite{deepfm, mmbattn, finalmlp, stec}. Deep CTR models utilize user features (e.g., age, gender), item features (e.g., category, price), and contextual features (e.g., weekday, hour) to infer users' interests in items. The primary components of a deep CTR model include the embedding layer, the feature interaction layer, and the output layer. The embedding and feature interaction layers are designed to capture interactions of varying orders between features represented as embeddings. Predictions are generated in the output layer and subsequently fed into a loss function. During the learning process, gradients are calculated using user-item interactions via back-propagation, and the resulting loss is minimized. The structure of the loss function is crucial in this optimization process.

CTR prediction inherently presents an asymmetric classification challenge, characterized by highly imbalanced class distributions. Positive class samples typically form coherent, compact distributions, whereas negative samples are more diverse. Cross-entropy loss is among the most widely used loss functions in deep learning \cite{deepfm, masknet, temporal}. Therefore, its binary variant, binary cross-entropy (BCE) loss with sigmoid activation, has been predominantly employed in deep CTR models. This loss function discriminates classes based on angular distances but tends to suffer from poor generalization due to overfitting in the presence of noisy data \cite{intro_R0}. Additionally, it is not tailored for asymmetric classification problems, failing to produce compact class acceptance regions for positive classes. Focal loss, on the other hand, offers greater robustness against class imbalance by introducing a dynamic weighting factor that depends on the error. This mechanism emphasizes hard-to-classify samples by penalizing their misclassifications more heavily. However, focal loss can be prone to the vanishing gradient problem, necessitating longer training durations \cite{intro_R2}. Bicici \cite{bicici} proposed the power cross-entropy (PCE) and power focal loss (PFL) methods by adding the power term to the these loss functions to get less loss from more confident samples and focus on harder samples. Hinge loss is also a common classification loss function that uses large margin principle to separate the positive classes from the negatives. However, similar to BCE loss, it is also not suited for asymmetric class classification problems.

All the loss functions summarized above (BCE, Focal and Hinge losses) do not attempt to return compact class acceptance regions by minimizing the intra-class variations. Additionally, as unbounded functions, they are particularly susceptible to noise \cite{intro_R3}.
There have been attempts to introduce loss functions that return compact class acceptance regions. To this end, Wen et al. \cite{intro_R4} integrated the softmax loss function with the center loss to improve face recognition. Center loss reduces the within-class variations by minimizing the distances between the individual face class samples and their corresponding class centers. 
Yang et al. \cite{R7} proposed Convolutional Prototype Network that learns several prototype vectors modeling each class. The class samples are enforced to compactly cluster in the vicinity of the learned prototype vectors, and novel test samples are rejected based on the distances to these prototype vectors. Similarly, Cevikalp et al. \cite{R8} approximated the class regions with compact hypersphere models.

As mentioned above, CTR prediction problem is an asymmetric problem where the few positive class samples are compactly clustered whereas the negative samples are much more and have diverse distributions. 
Therefore, we adopted a polyhedral conic deep neural network classifier in this study. This classifier uses the intersections of hyperplanes and cones to return compact, bounded polyhedral acceptance regions to approximate positive class samples. 
The utilized classification loss has also a compactness term that enforces the classifier to return more compact regions. It uses classical BCE loss for the inter-class separation.

Overall, the main contributions of this study can be summarized as follows:
\begin{itemize}
  \item We adopted a novel deep neural classifier employing a polyhedral conic functions for CTR prediction. This classifier addresses the CTR prediction problem as a one-class problem and incorporates a loss function specifically designed for asymmetric classification tasks.
  \item Due to its capacity to produce a compact acceptance region for positive samples (clicks), the developed method more accurately approximates the positive class samples, remaining unaffected by the number of classes.
  \item To validate the effectiveness of the proposed method, extensive large-scale experiments are conducted using different models across four distinct datasets. The experimental results indicate that the proposed method substantially enhances the baseline models utilizing binary cross-entropy (BCE) loss.
\end{itemize}

\begin{figure}[t]
  \centering
  \includegraphics[width=0.85\linewidth]{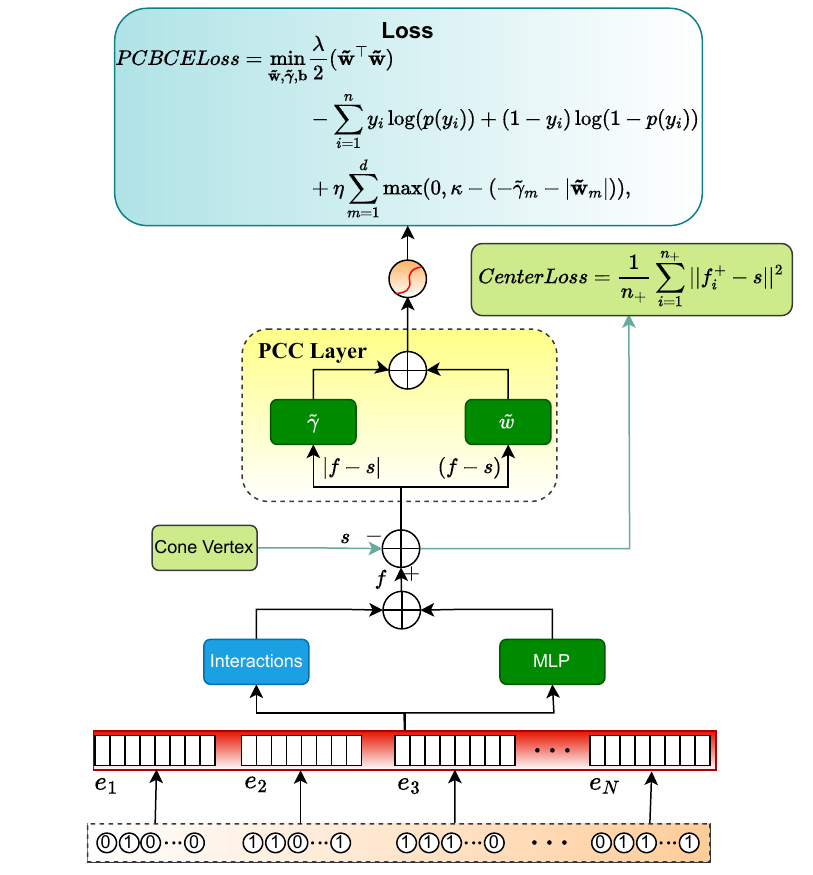}
  \caption{Demonstration of the proposed deep polyhedral conic classifier and proposed loss functions. In the training phase, the PCBCE loss updates all learnable weights except the cone vertex ($s$), while center loss term updates only the vertex of the cone.}
  \label{architecture}
\end{figure}

\section{Method}
General deep recommender system networks comprise three fundamental components: the embedding layer, the interaction/MLP layer, and the output layer. Figure \ref{architecture} provides an overview of the complete CTR prediction architecture, illustrating these layers in conjunction with the proposed layer and loss functions.
\subsection{Embedding Layer}

In CTR prediction models, high-dimensional and sparse input features, especially categorical variables, are common. Feature embedding layers capture relationships among these features. The embedding of a categorical variable \(\boldsymbol{x_i} \) is computed as \( \boldsymbol{e_i} = \boldsymbol{W_e} x_i \), where \( x_i \) is the one-hot representation of \( x \), a binary vector of size \( V \) (the vocabulary size). The embedding matrix \( \boldsymbol{W_e} \in \mathbb{R}^{V \times d} \) contains embedding vectors for each category. The embedding dimensionality \( d \) is a pre-set hyperparameter, yielding an embedding vector \(\boldsymbol{ e_i} \) of size \( \mathbb{R}^{1 \times d} \).

\subsection{Interaction and MLP Layers}
Once the embedding vectors are generated, they are subsequently processed through the interaction and multi-layer perceptron (MLP) layers. These layers facilitate the capturing and modeling of complex and non-linear relationships among the embedded features.



\subsection{Output Layer}
Following these layers, loss functions are employed to generate the prediction score ($\hat{y}$) and to minimize the discrepancies between the predicted scores and the actual labels.
We have used deep compact polyhedral conic classifier (DCPCC \cite{R5}) for classification. This classifier uses polyhedral conic functions that uses intersection of hyperplanes and $L1$ cones to discriminate relatively well localized positive classes from broader negative ones. It allows robust margin-based learning and returns convex, compact and bounded acceptance regions for positive classes when the learned classifier weights satisfy some pre-defined constraints. 

The Polyhedral Conic Functions \cite{R4} and Extended Polyhedral Conic Functions \cite{R3} respectively have the forms,
\begin{align}
 \!\! f_{\ww,\gamma,\cc,b}(\xx)&=\ww^\top(\xx-\cc)+\gamma\left\|\xx-\cc \right\|_1 -b, &\textrm{(PCF)}\!\!\label{eq:con_func} \\
  \!\! f_{\ww,\xgamma,\cc,b}(\xx)&=\ww^\top(\xx-\cc)+\xgamma^\top|\xx-\cc| -b, &\textrm{(EPCF)}\!\!\label{eq:econ_func}
\end{align}
In these equations, $\xx \in \R^d$ represents a test point, $\cc\in\R^d$ is the cone vertex, $\ww \in \R^d$ is a weight vector and $b$ is an offset. For PCF, $\left\|\uu\right\|_1=\sum_{i=1}^d|u_i|$ denotes the vector $L1$ norm and $\gamma$ is a corresponding weight, while for EPCF, $\left|\uu\right|=\left(|u_1|,\ldots,|u_d|\right)^\top$  denotes the component-wise modulus and $\xgamma$ is a corresponding weight vector. 

The polyhedral conic classifiers use functions of these forms, with decision regions $f(\xx) < 0$ for positives and $f(\xx) \geq 0$ for negatives. 
In both cases the positive region is essentially a hyperplane-section through an $L1$ cone centred at vertex $\cc$. 
It's worth noting that the acceptance regions of the classes become bounded and convex only under certain conditions: when both $b>0$, $\xgamma>0$, and when the absolute values of the weights $|w_i|$ are less than $\gamma_i$ for all $i=1,\ldots,d$. In simpler terms, this occurs when the slope of the hyperplane section is shallower than every facet of the $L1$ cone. Given that these conditions are satisfied, the resulting classifier returns compact and bounded region which is a deformed but still axis-aligned octahedral “kite” shape with overall size governed by $b$.

There are two factors that determine the hyper-volume of the class acceptance regions: 
$\gamma_i$ values and $b$. 
For fixed values of $\gamma_i$, as the $b$ value increases, the volume also increases because the vertex of the cone shifts to $(\cc,-b)$. This effectively moves the cone downward along the $L1$ axis without altering its facets. Conversely, decreasing the $b$ value reduces the volume as it shifts the cone upward towards the intersecting hyperplane. Similarly, for a constant  $b$ value, altering the volume is achievable by adjusting the slopes of the rays forming the cone through $\gamma_i$ values. Increasing $\gamma_i$ values diminishes the volume of the acceptance region, while decreasing them enlarges it. 

Let us assume that we are given labeled training samples in the form $(\xx_i,y_i)$, $i=1,\ldots,n$ and $y_i \in  \left\{-1,+1 \right\}$ in the binary classification problem. 
Let $\ff_i\in \R^d$ is the deep neural network feature representation of $\xx_i$ at the classification layer of the used network.
Consider that $\ww$ and $\xgamma$ are the weights of the extended polyhedral conic classifier used in the last stage of the deep neural network.
The authors of \cite{R5} set $\wn=-\ww$, $\gamman=-\xgamma$ (this is just for converting decision function into well-known SVM decision form), and then write the loss function of the deep neural network classifier as

\begin{equation}
\label{eq:mopt}
\underset{\wn, \gamman,\bb}{\text{min}} \:\:\:\:\: \frac{\lambda}{2} (\wn^\top \wn) + \sum_{i=1}^n \argmax(0, 1-y_i\left[\wn^\top (\ff_i-\cc) + \gamman^\top \left| \ff_i-\cc \right| + b \right] ) + \eta \sum_{m=1}^d \text{max}(0,\kappa-(-\gammani_{m}-\left|\wni_{m}\right|)),
\end{equation}

where $\lambda$ and $\eta$ are the weight parameters that must be set by the user. The first term in the loss function (\ref{eq:mopt}) is the regularization term that minimizes the $L2$ norm of the $\wn$ weight vector. The second loss term is the classical hinge loss function and it enforces to separate the positive and negative classes with a large margin. The last loss term controls the compactness of the positive class acceptance region. It simply uses a hinge loss term, and the positive $\kappa$ value acts as a margin to make sure that the positive $\gamma$ value is larger than $|w_i|$ by at least $\kappa$ in all feature directions, $m=1,\ldots,d$. This ensures that the positive class acceptance region is convex and bounded as described earlier. The classes become more tight and compact as the value of $\kappa$ increases.

In this study, we used deep compact polyhedral conic classifier with a small change. We utilized binary cross entropy loss function instead of classical hinge loss term used in the overall loss function given in (\ref{eq:mopt}). In this case, our loss function can be written as,

\begin{equation}
\label{eq:mopt2}
\underset{\wn, \gamman,\bb}{\text{min}} \:\:\:\:\: \frac{\lambda}{2} (\wn^\top \wn) - \sum_{i=1}^n y_i\log(p(y_i)) + (1-y_i)\log(1-p(y_i)) 
+ \eta \sum_{m=1}^d \text{max}(0,\kappa-(-\gammani_{m}-\left|\wni_{m}\right|)),
\end{equation}

where the predicted probability, $p(y_i)$, the example belongs to class 1 is computed by using,
\begin{equation}
p(y_i)  \:\: =  \:\: \sigma(\wn^\top (\ff_i-\cc) + \gamman^\top \left| \ff_i-\cc \right| + b),
\end{equation}
where $\sigma(z)=1/(1+e^{-z})$ is the logistic sigmoid function.

Cone vertex, $\cc$ must be set to the center of the positive samples. However, computing average of positive samples in each batch is computationally intractable. Therefore, we used center loss term to estimate cone vertex in each iteration. The center loss term minimizes the distances between the positive samples and  cone vertex $\cc$ as,
\begin{equation}
\underset{\cc}{\text{min}} \:\: \frac{1}{n_+}\sum_{i=1}^{n_+}\left\|\ff_i^+-\cc\right\|^2,
\end{equation}
where $\ff_i^+$ represents the samples coming from the positive class.
This loss term is just used to estimate the cone vertex and it is updated separately from the main network parameters.

\section{Experiments}
\subsection{Experimental Setup}

\textbf{Datasets}: The experiments are conducted on four widely used open benchmark CTR prediction datasets: Criteo, Avazu, MovieLens, and Frappe\footnote{\url{https://openbenchmark.github.io/BARS/}}. These datasets are acquired from BARS Benchmark \cite{BARS} which are preprocessed as Cheng et al. \cite{AFM}. The statistics for the used datasets are given in Table \ref{tab:datasets}.

\begin{table}[H]
\caption{The statistics of the datasets.}
\label{tab:datasets}
\centering
\resizebox{0.5\textwidth}{!}{%
\begin{tabular}{c c c}
\hline
\textbf{Dataset} & \textbf{Number of Samples} & \textbf{Number of Features} \\ \hline
Criteo           & 45,840,617                 & 39                          \\
Avazu            & 40,428,967                 & 22                          \\
MovieLens        & 2,006,859                  & 3                           \\
Frappe           & 288,609                    & 10                          \\ \hline
\end{tabular}%
}
\end{table}

\textbf{Baseline Models}: We conducted experiments on four baseline models. DNN \cite{youtubednn} uses multilayer perceptron, DCNv2 \cite{DCNv2} explicitly applies feature crossing as matrix to capture the feature interactions of different orders, AutoInt \cite{AutoInt} utilizes the multihead self-attention mechanism for the interaction layer, MaskNet \cite{masknet} introduces element-wise multiplication with instance-guided masks to capture high-order feature interactions.

\textbf{Evaluation Metrics}: Area under curve (AUC) and logloss are used as the evaluation metrics. It should be noted that 0.1\% increase in AUC is considered as significant for CTR prediction \cite{DCNv2, AutoInt, AFN, masknet}. 
Additionally, relative improvement (RelaImp) is used to show AUC improvements better \cite{masknet, CausalInt}. Since 50\% AUC is from randomness, we simply remove that constant part of AUC of both the proposed model and the base model to measure RelaImp.

\textbf{Implementation}: The proposed method is implemented in PyTorch 2.3.0 in FuxiCTR\footnote{\url{https://github.com/reczoo/FuxiCTR}\label{fuxictr}}, a reproducible CTR prediction benchmarking library \cite{BARS}. The implementation steps are followed for data preprocessing and data split, and seeding for a fair comparison. Baseline models are experimented with the reported best configurations in FuxiCTR. We optimized the model and the cone vertex separately using Adam and SGD optimizers, respectively, with a learning rate of 0.1 for the vertex optimizer and similar learning rates (0.0005-0.001) as BARS benchmark parameters for model optimizers. Reduce-lr-on-plateau is applied in both model and cone vertex optimizers. In general, baseline models are made less complex while integrating PCBCE to them to assure that the proposed method reduces model complexity. PCC layer is integrated to the baseline models as a layer before activation function. As in the benchmark configurations, embedding dimension is set to 10. Other hyperparameters are tuned for PCBCE.

\section{Experimental Results}

To investigate the effects of PCBCE, we conducted comprehensive experiments. Since it is designed as a plug-and-play method, we compare it with the baseline models’ individual performances. The results are shown in Table \ref{tab:results_general}.

\begin{table}[]
\centering
\caption{Comprehensive experimental results of PCBCE and baseline models on four open benchmark datasets.}
\label{tab:results_general}
\begin{tabular}{lcccclccc}
\multicolumn{4}{c}{\textbf{Avazu}}                                         &  & \multicolumn{4}{c}{\textbf{Criteo}}                                        \\ \cline{1-4} \cline{6-9} 
\multicolumn{1}{c}{} & \textbf{Base} & \textbf{PCBCE}   &                  &  & \multicolumn{1}{c}{} & \textbf{Base} & \textbf{PCBCE}   &                  \\ \cline{1-4} \cline{6-9} 
\multicolumn{1}{c}{} & \textbf{AUC}  & \textbf{AUC}     & \textbf{RelaImp} &  & \multicolumn{1}{c}{} & \textbf{AUC}  & \textbf{AUC}     & \textbf{RelaImp} \\ \cline{1-4} \cline{6-9} 
DNN                  & 0.76330       & \textbf{0.76464} & 0.51\%           &  & DNN                  & 0.81368       & \textbf{0.81379} & 0.04\%           \\
DCNv2                & 0.76364       & \textbf{0.76558} & 0.74\%           &  & DCNv2                & 0.81394       & \textbf{0.81429} & 0.11\%           \\
AutoInt              & 0.76237       & \textbf{0.76255} & 0.07\%           &  & AutoInt              & \textbf{0.81259}       & 0.81165          & -0.30\%          \\
MaskNet              & \textbf{0.76429}       & 0.76427          & -0.01\%          &  & MaskNet              & \textbf{0.81391}       & 0.81353          & -0.12\%          \\ \cline{1-4} \cline{6-9} 
\multicolumn{1}{c}{} &               &                  &                  &  & \multicolumn{1}{c}{} &               &                  &                  \\
\multicolumn{4}{c}{\textbf{Movielens}}                                     &  & \multicolumn{4}{c}{\textbf{Frappe}}                                        \\ \cline{1-4} \cline{6-9} 
\multicolumn{1}{c}{} & \textbf{Base} & \textbf{PCBCE}   &                  &  & \multicolumn{1}{c}{} & \textbf{Base} & \textbf{PCBCE}   &                  \\ \cline{1-4} \cline{6-9} 
\multicolumn{1}{c}{} & \textbf{AUC}  & \textbf{AUC}     & \textbf{RelaImp} &  & \multicolumn{1}{c}{} & \textbf{AUC}  & \textbf{AUC}     & \textbf{RelaImp} \\ \cline{1-4} \cline{6-9} 
DNN                  & 0.96768       & \textbf{0.96960} & 0.41\%           &  & DNN                  & 0.98405       & \textbf{0.98563} & 0.33\%           \\
DCNv2                & 0.96866       & \textbf{0.97081} & 0.46\%           &  & DCNv2                & 0.98382       & \textbf{0.98503} & 0.25\%           \\
AutoInt              & 0.96629       & \textbf{0.96743} & 0.24\%           &  & AutoInt              & 0.98309       & \textbf{0.98567} & 0.54\%           \\
MaskNet              & 0.96720       & \textbf{0.96872} & 0.33\%           &  & MaskNet              & 0.98368       & \textbf{0.98437} & 0.14\%           \\ \cline{1-4} \cline{6-9} 
\end{tabular}%
\end{table}

The evaluation on the Avazu dataset shows that PCBCE achieves a relative improvement (RelaImp) of 0.51\% for DNN, 0.74\% for DCNv2, and 0.07\% for AutoInt. MaskNet sees a slight decrease with a RelaImp of -0.01\%. Overall, PCBCE notably enhances performance in DNN and DCNv2 models.

The evaluation conducted on both the MovieLens and Frappe datasets collectively showcases the efficacy of PCBCE in enhancing predictive performance across diverse model architectures. Specifically, PCBCE yields relative improvements (RelaImp) of 0.41\% for DNN, 0.46\% for DCNv2, 0.24\% for AutoInt, and 0.33\% for MaskNet on the MovieLens dataset. Similarly, on the Frappe dataset, PCBCE achieves relative improvements of 0.33\% for DNN, 0.25\% for DCNv2, 0.54\% for AutoInt, and 0.14\% for MaskNet. These findings underscore the consistent and beneficial impact of PCBCE across different datasets and model structures.
On the Criteo dataset, the proposed loss function shows a relative improvement of 0.11\% for DCNv2 and minimal improvement for DNN. However, it leads to a relative underperformance of 0.30\% and 0.12\% for AutoInt and MaskNet methods respectively.

\subsection{Ablation Studies}
Table \ref{tab:results_ablation} presents the results of ablation studies conducted with DCNv2 on the MovieLens and Frappe datasets, examining the effects of the important designs of PCBCE. 


In the assessment using the MovieLens dataset, Hinge Loss emerges as the more effective option compared to BCE. Conversely, when considering the Frappe dataset, BCE outperforms Hinge Loss in terms of performance. The adapted loss function proposed by Cevikalp et al. \cite{R5}, termed \textbf{PCHinge} and represented by Eq. \ref{eq:mopt}, specifically tailored for CTR prediction, exhibits suboptimal performance when contrasted with both Hinge Loss and BCE across both datasets.

Our proposed method, PCBCE, consistently surpasses both BCE and Hinge Loss across both datasets. However, in the variant incorporating the $L1$ norm, denoted as \textbf{PCBCE\textsubscript{$L1$}} and defined by the PCF function in Eq. \ref{eq:econ_func}, it demonstrates a performance inferior to both BCE and Hinge Loss on the MovieLens dataset. Nonetheless, it achieves superior results compared to Hinge Loss and comparable outcomes to BCE on the Frappe dataset.

Finally, the simplified version of the proposed PCBCE method (deonted as \textbf{PCBCE ($\eta=0$)}), obtained by removing the compactness term by setting $\eta=0$ in Eq. \ref{eq:mopt2}, also exhibits superior performance relative to both BCE and Hinge Loss, particularly regarding the AUC metric.

\begin{table}[H]
\centering
\caption{Results for ablation studies. They are conducted with DCNv2 on MovieLens and Frappe datasets.}
\label{tab:results_ablation}
\begin{tabular}{lcc}
\cline{2-3}
                    & \multicolumn{1}{c|}{\textbf{MovieLens}} & \textbf{Frappe} \\ \cline{2-3} 
                    & \multicolumn{1}{c|}{\textbf{AUC (\%)}} & \textbf{AUC (\%)}  \\  \hline
BCE                 & \multicolumn{1}{c|}{96.866}             & 98.382          \\
Hinge               & \multicolumn{1}{c|}{96.891}             & 98.109          \\
PCHinge             & \multicolumn{1}{c|}{96.800}             & 98.051          \\ \hline
PCBCE\textsubscript{$L1$} & \multicolumn{1}{c|}{96.693}             & 98.356          \\ \hline
PCBCE ($\eta=0$)               & \multicolumn{1}{c|}{96.927}             & 98.437          \\ \hline
PCBCE               & \multicolumn{1}{c|}{97.004}             & 98.503          \\ \hline
\end{tabular}
\end{table}



\section{Conclusion}
In this paper, we introduce a novel deep neural network classifier utilizing polyhedral cone functions for click-through rate (CTR) estimation in industrial recommendation systems. This approach effectively addresses the challenges posed by numerical imbalance and geometric asymmetry in CTR prediction tasks. Our method, employing quasi-linear discriminants with distorted positive regions, focuses on creating compact, bounded acceptance regions for positive class examples necessary to improve prediction accuracy on imbalanced and asymmetric datasets.

Comprehensive experiments conducted on four general datasets demonstrate significant performance improvements of our proposed method over traditional approaches using Binary Cross-Entropy (BCE) loss on the Avazu, MovieLens, and Frappe datasets, while yielding varying results on Criteo with BCE across different models. The findings highlight how effective and efficient the polyhedral cone classifier is, especially in handling the inherent complexities of CTR prediction tasks.

\newpage
\bibliographystyle{unsrt}  
\bibliography{myreferences}

\end{document}